# Quantitative Decoding of Interactions in Tunable Nanomagnet Arrays Using First Order Reversal Curves

Dustin A. Gilbert,[1] Gergely T. Zimanyi,[1,*] Randy K. Dumas,[1] Michael Winklhofer,[2] Alicia Gomez,[3] Nasim Eibagi,[1] J. L. Vicent,[3,4] and Kai Liu[1,*]

[1]*Dept. of Physics, University of California, Davis, California, 95616, USA*
[2]*Dept. of Earth & Environmental Sciences, Ludwig-Maximilians-Universität München, Germany*
[3]*Dept. Fisica Materiales, Universidad Complutense, 28040 Madrid, Spain*
[4]*IMDEA-Nanociencia, Cantoblanco 28049, Madrid, Spain*

[*]Corresponding authors. Email: (G.T.Z.) zimanyi@physics.ucdavis.edu; (K.L.) kailiu@ucdavis.edu.

**Abstract**

To develop a full understanding of interactions in nanomagnet arrays is a persistent challenge, critically impacting their technological acceptance. This paper reports the experimental, numerical and analytical investigation of interactions in arrays of Co nanoellipses using the first-order reversal curve (FORC) technique. A mean-field analysis has revealed the physical mechanisms giving rise to all of the observed features: a shift of the non-interacting FORC-ridge at the low-$H_C$ end off the local coercivity $H_C$ axis; a stretch of the FORC-ridge at the high-$H_C$ end without shifting it off the $H_C$ axis; and a formation of a tilted edge connected to the ridge at the low-$H_C$ end. Changing from flat to Gaussian coercivity distribution produces a negative feature, bends the ridge, and broadens the edge. Finally, nearest neighbor interactions segment the FORC-ridge. These results demonstrate that the FORC approach provides a comprehensive framework to qualitatively and quantitatively decode interactions in nanomagnet arrays.

Nanomagnet arrays are basic building blocks[1] for key technologies such as ultrahigh density magnetic recording media,[2-4] magnetic random access memory (MRAM),[5,6] and logic devices.[7-9] Interactions within the arrays critically affect the functionalities of the nanomagnets as well as enable new device concepts. For instance, dipolar interactions may trigger an analog memory effect in nanowire arrays,[10] enable digital computation in magnetic quantum-dot cellular automata systems,[7,8,11] lead to frustrations in artificial "spin ice",[12-14] or adversely affect thermal stability and switching field distribution in magnetic recording media or MRAM elements.[5,15-17] Probing and managing these interactions is often difficult because they are long-ranged, anisotropic, and configuration-dependent.[17]

The first-order reversal curve (FORC) method[18,19] has provided detailed characterization for a variety of magnetic[20-27] and other hysteretic systems.[28,29] However, a coherent framework to interpret the features of the FORC diagrams and extract quantitative information is still lacking despite decades of effort by numerous groups. In this work, using the FORC method we have quantitatively investigated tunable interactions in model systems of single domain nanomagnet arrays. With mean-field level simulations, supplemented with a cluster extension, we have reproduced *all* features and trends of the experimental FORC diagrams quantitatively and identified their physical origins. Our approach decodes interactions in nanomagnet arrays, even disordered arrays, and also presents a pathway to evaluate (de)stabilizing interactions in other hysteretic systems.

**Results**

Rectangular arrays of polycrystalline Co ellipses were fabricated with varying center-to-center spacing by magnetron sputtering, in conjunction with e-beam lithography and lift-off.

Details are presented in Methods. The ellipses have major/minor axes of 220/110 nm, with a structure of Ta(1nm)/Co(9nm)/Ta(1nm), forming 50×50 μm$^2$ arrays. In arrays A1/2/3 the minor-axis spacings of 150/200/250 nm are less than the major-axis spacing of 500 nm. Thus the mean dipolar interactions are demagnetizing, favoring anti-parallel alignment. In arrays B1/2/3 the minor-axis spacing of 500 nm exceeds the major-axis spacings of 250/300/350 nm. Therefore, the mean dipolar interactions are magnetizing, favoring parallel alignment. Scanning electron microscopy (SEM) and magnetic force microscopy (MFM) images, at remanence after DC demagnetization, of arrays A1 and B1 are shown in Figs. 1 and 2, respectively. MFM image contrast indicates the out-of-plane stray fields and confirms the ellipses' single domain state.

FORC measurements were performed to obtain magnetization $M(H,H_R)$ under different reversal field $H_R$ and applied field $H$. The FORC-distribution is then extracted,[18] $\rho(H, H_R) \equiv -\frac{1}{2M_S}\frac{\partial^2 M(H,H_R)}{\partial H\, \partial H_R}$, where $M_S$ is the saturation magnetization. In loose analogy to the Preisach model, the FORC-distribution in certain simple cases can be interpreted as the 2-dimensional distribution of elemental hysteresis loops with unit magnetization called hysterons on the $(H,H_R)$ plane, or on the corresponding $(H_C,H_B)$ plane, defined by local coercivity $H_C=(H-H_R)/2$ and bias/interaction field $H_B=(H+H_R)/2$.

Good agreement between measured and simulated FORC-distributions was obtained for all studied arrays [e.g., Figs. 1(c) and (d) for array A2]. All demagnetizing FORCs A1/2/3 exhibited a ridge with the high-$H_C$ end on the $H_C$ axis and the low-$H_C$ end shifted in the $+H_B$ direction.[30] Increasing interactions in A3→A2→A1 increased the low-$H_C$ shift and the length of the ridge. In addition, an edge emerged from the low-$H_C$ end towards negative $H_B$, highlighted by the arrows, forming a "wishbone" or boomerang structure.[24] A negative feature is observed at negative $H_B$ values near the high-$H_C$ end. Similarly, FORC distributions for the magnetizing arrays also exhibit a ridge with the high-$H_C$ end on the $H_C$ axis, but with the low-$H_C$ end shifted towards -$H_B$ [e.g., Figs. 2(c) and (d) for array B2]. Increasing interactions again increase the low-$H_C$ shift, but reduce the length of the ridge.[30] A negative feature below the ridge is more prominent than that in the demagnetizing case.

Non-interacting Case: A FORC gives a non-zero contribution to $\rho(H, H_R)$ only if $dM/dH$ along the FORC depends on $H_R$. We first show that in the non-interacting case $\rho$ coincides with the coercivity distribution $D(H_K)$, spread along the $H_C$ axis. Indeed, particle $P_i$ with coercivity $H_K^i$ down-flips at $H_{dn}^i=-H_K^i$ and up-flips at $H_{up}^i=H_K^i$. Therefore, on a FORC starting at $H_R> -H_K^i$, $P_i$ starts out up-flipped and remains so, not contributing to $dM/dH$ nor $\rho$.

In contrast, on FORC($H, H_R=-H_K^i$), $P_i$ is the *last to down-flip* along the major loop, and has the highest coercivity among the down-flipped particles. Therefore, $P_i$ is the *last to up-flip* as $H$ increases past $H_K^i$ on the same FORC, causing a $dM/dH>0$ jump. Since this $dM/dH$ jump is *unmatched* by the neighboring FORC($H,H_R>-H_K^i$), $dM/dH$ exhibits a dependence on $H_R$, making $d(dM/dH)/dH_R$ non-zero. $dM/dH$ increases as $H_R$ decreases, making a positive contribution to $\rho$ at $(H=H_K^i,H_R=-H_K^i)$.

For all subsequent FORCs at $H_R<-H_K^i$, $P_i$ starts out down-flipped but still up-flips at $H=H_K^i$. The $dM/dH$ jumps on these FORCs are *matched* since they occur at the *same field* on each FORC($H,H_R<-H_K^i$). Thus $dM/dH$ is independent of $H_R$, and doesn't contribute to $\rho$. This reasoning highlights that only $dM/dH$ jumps on individual FORCs that are *unmatched* by neighboring FORCs contribute to $\rho$. Each particle $P_i$ contributes to $\rho$ only once, at $(H=H_K^i, H_R=-H_K^i)$ or equivalently at $(H_C=H_K^i, H_B=0)$. The contributions of all particles gives rise to a ridge along the $H_C$ ($H_R=-H$) axis, which reflects $D(H_K)$.

Interacting Case: Next, we introduce interactions between nanomagnets on the mean-field level by including an interaction field $H_{int}=\alpha M(H)$, where $\alpha<0$ for demagnetizing systems and $\alpha>0$ for magnetizing ones.[21] Fig. 3(a) shows a sequence of FORCs for a demagnetizing system with a rectangular $D(H_K)$, and a zoom-in view of the boxed region (right panel). The three FORC segment-pairs (1)/(2)/(3) show that the last $dM/dH$ jump on each FORC$(H,H_R=-H_K^i-\alpha M(H_R))$ - caused by the last up-flipping particle $P(H_K^i)$ - is *unmatched* by the neighboring FORC$(H,H_R>-H_K^i-\alpha M(H_R))$.

Fig. 3(b) shows that with interactions the unmatched $dM/dH$ jumps still generate the ridge, but at shifted $H_B$ values. Importantly, on the mean-field level all particles experience the same interaction field and thus the order of flips continues to be governed by the order of the coercivities: along the major loop the particles down-flip in the order of their coercivities, lowest (highest) coercivity particle $P(H_K^{min})$ first [$P(H_K^{max})$ last]. Starting at the low-$H_C$ end, the lowest coercivity particle $P(H_K^{min})$ down-flips at $H_{dn}^{min}=-H_K^{min}-\alpha M_S$, as no other particles have flipped yet: $M(H_{dn}^{min})=M_S$. Increasing $H$ along FORC$(H,H_R=H_{dn}^{min})$, $P(H_K^{min})$ up-flips at $H_{up}^{min}=H_K^{min}-\alpha M_S$, causing a positive jump $dM/dH>0$ as shown by the *lower* FORC$(H,H_R=H_{dn}^{min})$-segment of pair (1). This jump, caused by $P(H_K^{min})$, is *absent* on the *upper* FORC$(H,H_R>H_{dn}^{min})$-segment and is thus *unmatched*, contributing to $\rho$ at $(H=H_K^{min}-\alpha M_S, H_R=-H_K^{min}-\alpha M_S)$, or similarly at $(H_C=H_K^{min}, H_B=-\alpha M_S)$, defining the low-$H_C$ end. These flipping fields are shifted from their non-interacting values, as shown by the arrow set (1) in Fig. 3(b). Since $P(H_K^{min})$ defines the low-$H_C$ end of the FORC-ridge, one concludes that interactions shift the low-$H_C$ end of the FORC-ridge to the $+H_B$ direction by $-\alpha M_S$ (recall $\alpha<0$), but leave its $H_C$ coordinate un-shifted at $H_C=H_K^{min}$.

FORC-segment pairs (2)/(3) in Fig. 3(a) illustrate the up-flip of higher coercivity particles $P(H_K^i)$. $P(H_K^i)$s create unmatched $dM/dH$ jumps on FORCs where they were the last to down-flip at $H_{dn}=-H_K^i-\alpha M(H_R)$ [vertical arrows in Fig. 3(b)] and also are the last to up-flip at $H_{up}=H_K^i-\alpha M_S$ [horizontal arrows in Fig. 3(b)], since $M(H_{up}^i)=M_S$ when $P(H_K^i)$ up-flips.

The high-$H_C$ end of the FORC-ridge is defined by $P(H_K^{max})$ that is the last to down-flip when the rest of the system is already *negatively* saturated ($M=-M_S$), thus $H_{dn}^{max}=-H_K^{max}-(-\alpha M_S)$. $P(H_K^{max})$ up-flips along the FORC($H,H_R=H_{dn}^{max}$) only after the rest of the system is *positively* saturated: $H_{up}^{max}=H_K^{max}-\alpha M_S$. Accordingly, the *unmatched dM/dH* jumps caused by $P(H_K^{max})$ contributes to $\rho$ only at ($H=H_K^{max}-\alpha M_S, H_R=-H_K^{max}+\alpha M_S$), or similarly at ($H_C=H_K^{max}-\alpha M_S, H_B=0$), defining the high-$H_C$ end. As observed before, the high-$H_C$ end of the FORC-ridge remains *on* the $H_C$ axis,[24] but stretched along the $H_C$ axis by $-\alpha M_S$ ($\alpha<0$).

Note that interactions shift the FORC ridge feature in *H uniformly* [Fig. 3(b)], i.e., the resultant projection of FORC distribution onto the *H*-axis is only displaced from its intrinsic values, but not distorted. This *H*-projection therefore mirrors the non-interacting case, where $H^{Up}=H_C=H$, reflecting the intrinsic coercivity distribution, simply displaced by $-\alpha M_S$. Thus the intrinsic coercivity distribution can be - without distortion from interactions – directly identified from the FORC distribution.

Figs. 1(c,d) and 3(b,c) show that besides the ridge, $\rho$ exhibit an edge as well with interactions.[27] As discussed earlier, in the absence of interactions, the *dM/dH* jumps along a FORC($H,H_R=-H_K^i$) are *matched* by the jumps on the subsequent FORC($H,H_R<-H_K^i$)s, not contributing to $\rho$. The arrows of Fig. 3(a) show that the interactions destroy this matching specifically *at the low-$H_C$ end* by shifting the *first* up-flip field $H_{up}^{min}$ of each FORC, caused by $P(H_K^{min})$, by $-\alpha M(H_R)$. These shifts make the *dM/dH* jumps misaligned, see FORC-segment-pairs (4) and (5) (above and below), thus contributing to $\rho$ at ($H=H_K^{min}-\alpha M(H_R),H_R$). These unmatched jumps give rise to the edge in Fig. 3(c). The end-points of the edge are ($H=H_K^{min}-$

$\alpha M_S, H_R=-H_K^{min}-\alpha M_S)$ and $(H=H_K^{min}+\alpha M_S, H_R=-H_K^{max}+\alpha M_S)$, or alternatively $(H_C=H_K^{min}, H_B=-\alpha M_S)$ and $(H_C=(H_K^{min}+H_K^{max})/2, H_B=\alpha M_S+(H_K^{min}-H_K^{max})/2)$. Accordingly, the tilt and asymmetry of the edge provide a direct measure of the width of coercivity distribution $D(H_K)=H_K^{max}-H_K^{min}$. In the extreme case of nearly identical nanomagnets with $D(H_K)=H_K^{max}-H_K^{min}\approx 0$, the edge is vertical in the $H_C - H_B$ plane, at $H_C=H_K$ and $H_B$ within $\pm \alpha M_S$, as observed experimentally in Ni nanowire arrays.[10]

In short, on the mean-field level the FORC-distribution of a system with demagnetizing interactions exhibits an edge and a ridge, shifted by the unmatched first and last *dM/dH* jumps along each FORC. The FORC-distribution vanishes between them for the considered flat coercivity distribution, because jumps between the first and the last jumps along each FORC are matched by jumps on the neighboring FORCs. Here, the matched jumps are not caused by the same particles, as in the non-interacting system, but rather by different particles whose up-flipping fields were shifted into alignment by the interactions. Still, the jumps are *matched* because the values of the aligned jumps are the same on neighboring FORCs *for a flat distribution D(H_K)*. Visibly, the flat coercivity distribution on the mean-field already reproduces most features of the measured FORC-distribution.

To improve our model we introduce a more realistic Gaussian $D(H_K)$ to elucidate the origin of the negative features, which represent a clear distinction between FORC and a literal Preisach interpretation. A Gaussian breaks the matching of jumps as now shown by examining the set of particles $\{P(H_K^{Cent})\}$ around the center of the coercivity distribution. A $P(H_K^{Cent})$ particle is the *last* to down-flip at $H_R=-H_K^{Cent}$, where $M(H_R)=0$, and the *last* to up-flip on the FORC$(H, H_R=-H_K^{Cent})$, contributing to $\rho$ at $(H=H_K^{Cent}-\alpha M_S, H_R=-H_K^{Cent})$. On FORC$(H, H_R<-H_K^{Cent})$s $P(H_K^{Cent})$ is no longer the last to up-flip. On subsequent FORCs, up-flip jumps from different particles get shifted into alignment with this jump. For the flat distribution, the number of particles shifted into alignment is steady, making the match complete and thus zero contribution to $\rho$. However, for the Gaussian distribution, the particles shifted into alignment come from the *decreasing* slope of the Gaussian, leading to *dM/dH* jumps with a decreasing magnitude, providing only a *partial-match* and generating a *negative* contribution to $\rho$. Fig. 3(d) shows the FORC-distribution for a Gaussian model that indeed develops a negative region specifically tracking the decreasing slope of the Gaussian, highlighted by the dashed line. Analogous arguments show that a Gaussian $D(H_K)$ also bends the ridge and broadens the edge.

Nearest-neighbor correlations: The last unaccounted feature of the measured $\rho$ is the segmenting of the FORC-ridge into separate low-$H_C$ and high-$H_C$ ends, with different amplitudes. To explain this we include the nearest-neighbor interaction fields $H_{nn}^i(conf)$ as the first terms of a systematic cluster-expansion.

Decreasing the field from positive saturation, the weakest coercivity particles down-flip first. For demagnetizing interactions, $H_{nn}^i(conf)$ of these down-flipped particles stabilize their nearest-neighbors in their up state. Therefore, for a sufficiently narrow $D(H_K)$, the magnetization decreases towards zero by developing a checkerboard pattern [Fig. 1(b)]. The checkerboard naturally forms defects where the sequence of increasing coercivities selects third-nearest neighbor particles to down-flip. Still, the dominant reversal mechanism for nearly half of the particles is the checkerboard formation: down-flipping with all-up neighbors. Accordingly, the (nearly) half of the FORC ridge with $H_K^i < H_K^{Cent}$ gets shifted along the $+H_B$ axis by $H_{nn}^i(up)$, where $H_{nn}^i(up)$ is the interaction field for the all-neighbors-up configuration. Once the checkerboard pattern is formed, the rest of the particles flip with neighbors in various intermediate configurations. Therefore, the $H_K^i > H_K^{Cent}$ half of the ridge is broken into several pieces, shifted by varying $H_{nn}{}^i(conf)$ fields. Consequently, the nearest neighbor interactions manifest as segmenting of the ridge.[30]

To reiterate, the demagnetizing interactions ($\alpha<0$) (1) shift the non-interacting FORC-ridge at the low-$H_C$ end to the +$H_B$ direction by -$\alpha M_S$; (2) stretch the non-interacting FORC-ridge at the high-$H_C$ end along $H_C$ by -$\alpha M_S$ without shifting it off the $H_C$ axis; and (3) form a tilted edge connected to the ridge at the low-$H_C$ end. Changing from flat to Gaussian $D(H_K)$ distribution (4) produces a negative feature, bends the ridge, and broadens the edge. Finally, (5) nearest neighbor interactions segment the FORC-ridge.

<u>Magnetizing interactions</u>: Adapting the above arguments for magnetizing interactions ($\alpha>0$): (1) the low-$H_C$ end is shifted in the -$H_B$ direction, and (2) the high-$H_C$ end is compressed without shifting it off the $H_C$ axis [Fig. 3(f)]. (3) Regarding the edge, the first up-flip along each branch is shifted by interactions in the *opposite* direction as the demagnetizing case [Fig. 3(e) right panel]. Therefore, the first $dM/dH$ jumps are unmatched, decreasing in magnitude with *more negative $H_R$*, thus *negatively* incrementing the FORC, forming a *negative* edge [Fig. 3(g)]. Changing from flat to Gaussian $D(H_K)$ distribution (4) the negative edge gets pressed towards the positive ridge, and the FORC-ridge becomes curved [Fig. 3(h)]. The inclusion of nearest neighbor terms leads to (5) an avalanche reversal, collapsing the FORC-ridge to a single-value.[31]

<u>Quantifying Interaction Fields:</u> Finally, we demonstrate the quantitative predicting power of the above considerations. The lowest $H_K{}^i$, which is shifted in $H_B$ by $\alpha M_S$, is extracted from the FORC ridge by selecting an $H_C(threshold)$ such that 10% of the particles have $H_K{}^i<H_C(threshold)$, and averaging $\rho$ over the $H_C=0 \rightarrow H_C(threshold)$ range. The averaged $(dM/dH_B)'$ are shown in insets of Figs. 4(a) and 4(b). The $H_B$ shift is determined by linearly extrapolating $(dM/dH_B)'$ at the high |$H_B$| end to zero. The interaction field is calculated by a

finite element method (using the NIST OOMMF code) for the nearest and next nearest neighbors, and treating the remainder of the array as point dipoles. The experimental $H_B$ shifts and the calculated interaction fields agree remarkably well (Fig. 4), confirming the validity of the mean-field description of the FORC-distribution and its quantitative predictive power, making the FORC technique a tool to extract numerical values of interaction fields. This is particularly important for disordered arrays where calculations of interactions are not easily achievable.

**Discussion**

In this work, systems of interacting nanomagnets were examined experimentally, numerically, and analytically, using the FORC technique. A mean-field analysis based on the concept of unmatched jumps accounted for all experimentally observed features of the FORC diagram, including its shifted ridge-and-edge structure and negative features. The tilting, shifting, and stretching of these structures were identified as tools to extract quantitative information about the system, demonstrating the predictive power of the FORC technique. Construction of the FORC distribution through unmatched jumps, and recognizing the (de)magnetizing interactions as a particular case of (de)stabilizing interactions, presents an approach which can be used to evaluate any hysteretic system with the FORC technique.

**Methods**

Arrays of Co ellipses were fabricated by DC magnetron sputtering, in a vacuum chamber with a base pressure of $1 \times 10^{-8}$ Torr and Ar sputtering pressure of $2 \times 10^{-3}$ Torr, on Si (100) substrates, in conjunction with electron beam lithography and lift-off techniques. Magnetic hysteresis loops were measured at room temperature using the magneto-optical Kerr effect (MOKE) magnetometer with a 632 nm HeNe laser having a 30 μm spot-size, capturing the reversal behavior of ~5,000 ellipses.[32] The magnetic field was applied parallel to the major axis of the ellipses. Each measurement was averaged over ~$10^3$ cycles at a rate of 11 Hz. The arrays were coated with a 60 nm ZnS layer to improve the signal-to-noise ratio.[33] FORC measurements were performed as follows:[22] from positive saturation the magnetic field is swept to a reversal field $H_R$, where the magnetization $M(H,H_R)$ is measured under increasing applied field $H$ back to saturation, tracing out a FORC. The process is repeated for decreasing reversal field $H_R$.[30]

Ellipses were modeled as dipoles oriented parallel to their major axes. The inter-dipole spacing and magnetic moment per dipole in the 100×100 array were representative of the experimental system. Each dipole $i$ was assigned an intrinsic coercivity $H_K^i$ with a distribution experimentally determined from the sample having the weakest interactions, A3. The $H_{int}^i$ dipolar interaction fields at dipole $i$ were calculated on the mean-field level as $\alpha M(H)$, where $\alpha$ was calibrated such that $\alpha M_S$ equals the analytically calculated $H_{int}$ at saturation. This mean-field formulation was extended by the first term of a cluster expansion, representing the nearest neighbor dipole interaction $H_{nn}^i$ explicitly: $H_{int}^i = \alpha M(H) + H_{nn}^i$. At each field step ($\Delta H=1$Oe) the total field $H_{tot}^i = H + H_{int}^i$ was compared to $H_K^i$, down-flipping occurred when $H + H_{int}^i < -H_K^i$ and up-flipping occurred when $H + H_{int}^i > H_K^i$, until all dipoles became stable.

**Acknowledgement**

This work was supported by NSF (ECCS-0925626, DMR-1008791, ECCS-1232275) and BaCaTec (A4-[2012-2]). Work at UCM and IMDEA was supported by the Spanish MINECO grant FIS2008-06249 and CAM grant S2009/MAT-1726.

**Author Contributions**

D.A.G. obtained the experimental and simulation results, and wrote the first draft of the paper. G.T.Z. and M.W. participated in the simulation design. R.K.D., A.G., N.E., J.L.V. and K.L. participated in the experimental design, fabrication and characterization. K.L. and G.T.Z. designed and coordinated the whole project. All authors contributed to analysis, discussion and revision of the paper.

**Competing Financial Interests**

The authors declare no competing financial interests.

**Figure Captions:**

**Figure 1**. Demagnetizing Arrays: (a) SEM and (b) MFM image of the DC-demagnetized A1 array. Dashed ovals outline single ellipses, while the dashed box outlines an example of the checkerboard pattern. (c) Experimental and (d) simulated FORC distributions for the A2 array.

**Figure 2**. Magnetizing Arrays: (a) SEM and (b) MFM image of the DC-demagnetized B1 array. Dashed ovals outline single ellipses. (c) Experimental and (d) simulated FORC distributions for the B2 array.

**Figure 3**. (a) Schematic illustration of family of FORCs for arrays with a flat coercivity distribution and mean-field demagnetizing interactions, with bold lines and numbers indicating unmatched d$M$/d$H$ jumps. Calculated FORC distributions are shown in (b) illustrating the construction of the ridge and (c) the edge. (d) FORC distribution with the same interactions, but a Gaussian coercivity distribution; emergent negative feature is indicated by the dashed boundary. Similar panels are shown in (e-h) for the magnetizing case.

**Figure 4**. Calculated (open symbols) and experimentally determined (solid symbols) interaction field for (a) demagnetizing arrays A1/2/3 and (b) magnetizing arrays B1/2/3. Averaged FORC-distribution utilizing the $H_C(threshold)$ are shown in insets for the (a) A2 and (b) B2 array, where the linear extrapolation is illustrated by the dashed line and open circle.

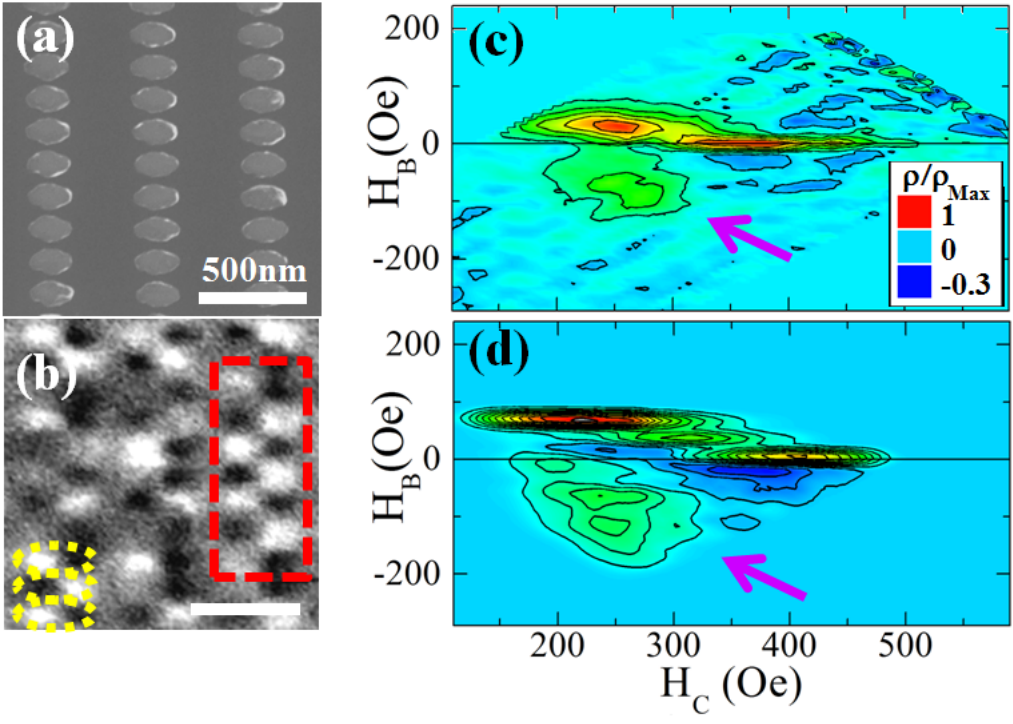

**Fig. 1.**

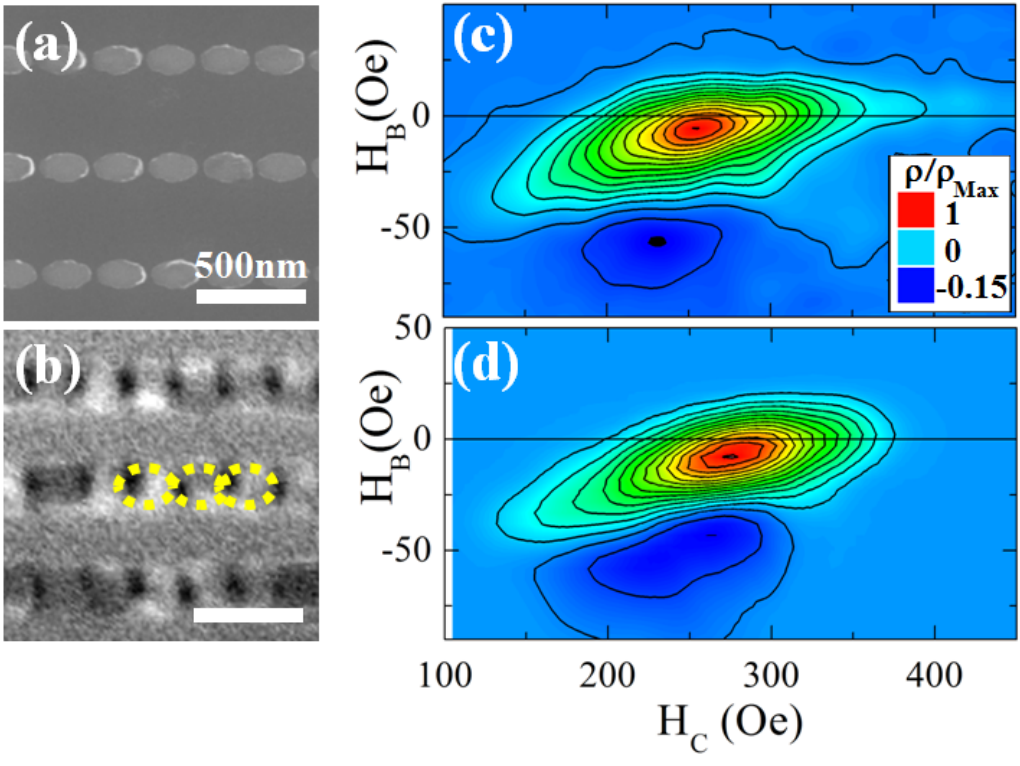

**Fig. 2.**

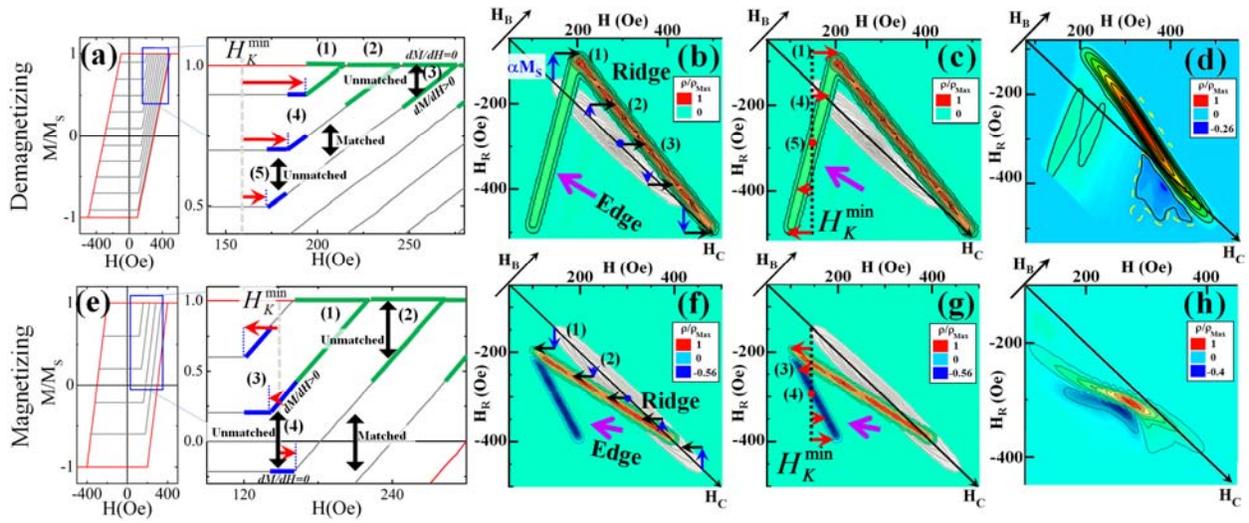

**Fig. 3, Gilbert et. al.**

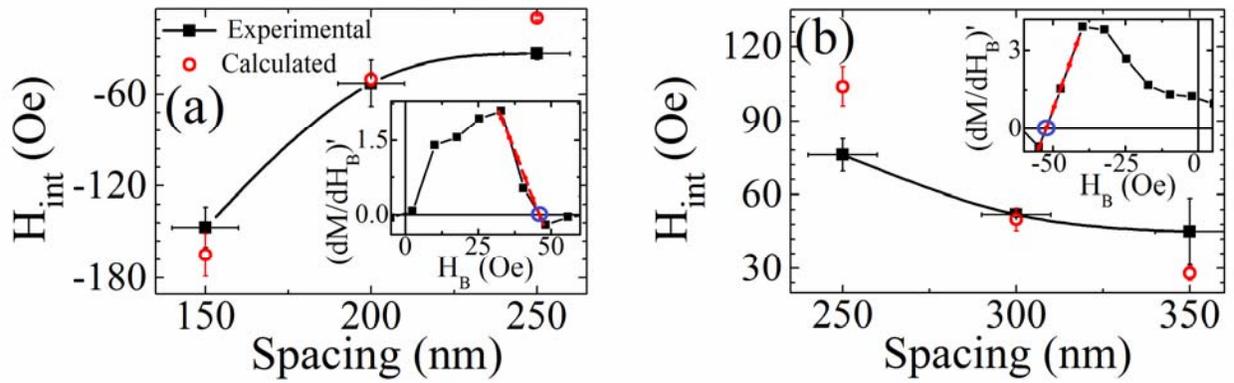

**Fig. 4. Gilbert et. al.**

**Quantitative Decoding of Interactions in Tunable Nanomagnet Arrays**

**Using First Order Reversal Curves**

Dustin A. Gilbert,[1] Gergely T. Zimanyi,[1] Randy K. Dumas,[1] Michael Winklhofer,[2]

Alicia Gomez,[3] Nasim Eibagi,[1] J. L. Vicent,[3,4] and Kai Liu[1]

[1]Dept. of Physics, University of California, Davis, California, 95616, USA
[2]Dept. of Earth & Environmental Sciences, Ludwig-Maximilians-Universität München, Germany
[3]Dept. Fisica Materiales, Universidad Complutense, 28040 Madrid, Spain
[4]IMDEA-Nanociencia, Cantoblanco 28049, Madrid, Spain

**Supplemental Material**

<u>Experiments-Demagnetizing arrays:</u> The physics of interactions were probed with the FORC technique by measuring six arrays where the interactions were systematically tuned. Two of the measured FORC diagrams were highlighted in Figs. 1 and 2 in the main text. This Supplemental material presents the full set, to demonstrate the experimental trends. Figure S1 shows the experimentally determined family of FORCs, the corresponding measured FORC distribution and the simulated FORC distribution. Here the demagnetizing interactions are strongest for array A1 (left column) and weakest for A3 (right column).

The families of FORCs become increasingly sheared with increasing interactions (top row, A3→A2→A1). Shearing of the hysteresis loops has been previously observed and is caused by mean-field demagnetizing interactions. The experimental FORC distributions exhibit a ridge aligned with the local coercivity $H_C$ axis, shifted in the $+H_B$ direction at the low-$H_C$ end, the shift increasing with increasing interaction strength (center and bottom rows, right to left). The high-$H_C$ end of the FORC ridge remains on the $H_C$ axis ($H_B=0$), while the extent or length of the ridge increases with increasing interaction strength, stretching the ridge.

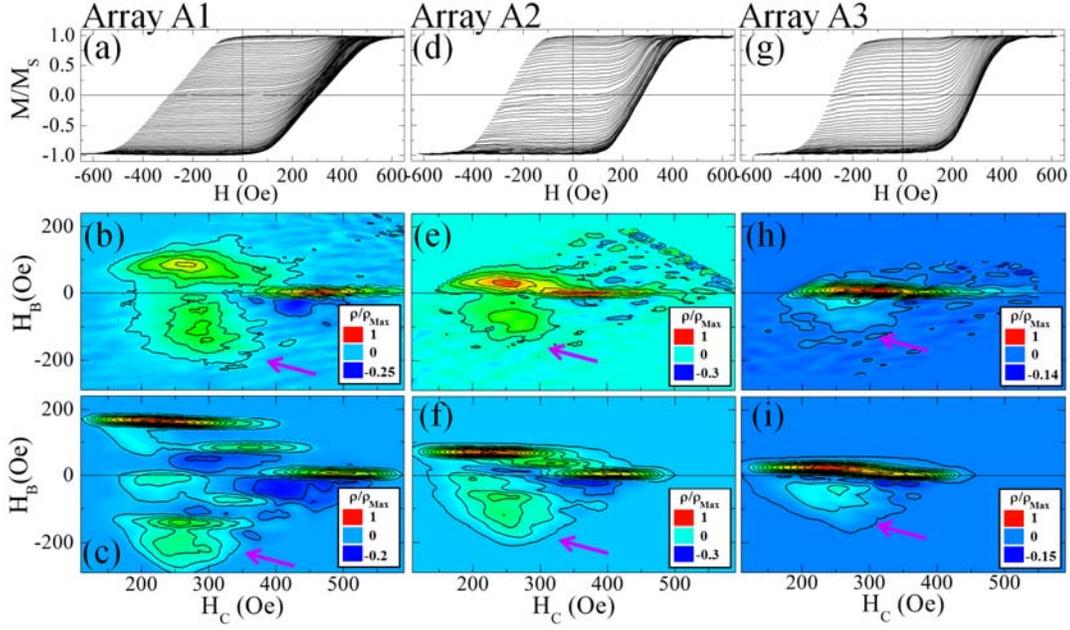

*Fig. S1: Experimentally determined family of FORCs (top row) and FORC distributions (middle row), and simulated FORC distributions (bottom row) for array A1 (panels a-c), A2 (panels d-f), and A3 (panels g-i). Vertical edge is identified by the purple arrow.*

In addition, an edge directed along the $-H_B$ direction develops, highlighted by the arrow in Fig. S1, attached to the ridge at the low-$H_C$ end. The edge becomes more extended with stronger interactions. A negative feature also develops underneath the high-$H_C$ end. All these trends are consistent with the simulations and the predictions of the mean-field theory.

In addition, in Fig. S1 both the experiments and the simulations show that the ridge and the edge are not smooth but segmented. Segmenting is caused by the nearest neighbor interactions, as explained in the main text and in relation to Fig. S3 below.

<u>Experiments-Magnetizing arrays</u>: FORC diagrams for the magnetizing arrays B1/2/3 are shown in Fig. S2. The family of FORCs becomes *less* sheared (more square) with increasing interactions (top row, B3→B2→B1), in contrast to the demagnetizing case. The FORC distributions exhibit a ridge aligned with the $H_C$ axis, shifted in the $-H_B$ direction at the low-$H_C$ end. The shift increases with increasing interaction strength (center and bottom rows, right to left). The high-$H_C$ end of the FORC ridge remains on the $H_C$ axis ($H_B=0$), while the extent of the ridge along $H_C$ decreases with increasing interaction strength – the ridge becomes compressed. Lastly, a negative edge develops below the ridge. Similarly to Fig. S1, all trends of the experimental results are reproduced in the simulated results.

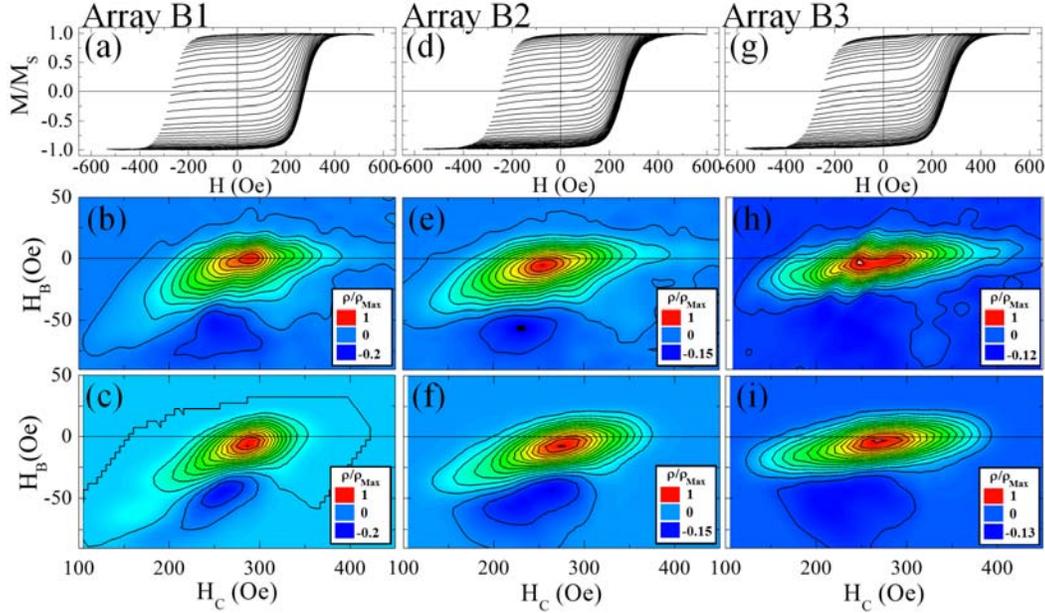

*Fig. S2: Experimentally determined family of FORCs (top row) and FORC distributions (middle row), as well as simulated FORC distributions (bottom row) for array B1 (panels a-c), B2 (panels d-f), and B3 (panels g-i).*

The features and trends exhibited by the magnetizing arrays are explained next based on the mean field theory of the FORC diagrams, analogous to the demagnetizing interactions in the main text. Specifically, on each FORC curve, the highest coercivity down-flipped particle was the last to down-flip at $H_{dn}=-H_K^i-\alpha M(H_R)$, since it experienced a mean field $\alpha M(H_R)$.

The same highest coercivity particle on each FORC curve is the last to up-flip at $H_{up}=H_K^i-\alpha M_S$, as the array reaches full saturation. As for the demagnetizing case, the corresponding last $dM/dH$ jump on each FORC($H$, $H_R=-H_K^i-\alpha M(H_R)$) is *unmatched* by the neighboring FORC($H, H_R>-H_K^i-\alpha M(H_R)$) which differs only in that $P(H_K^i)$ was un-flipped along that FORC.

These unmatched *dM/dH* jumps generate the FORC-ridge aligned with the $H_C$ axis.

In complete analogy to the demagnetizing case, the ridge is shifted off the $H_C$ axis at the low-$H_C$ end by $-\alpha M_S$, but because $\alpha>0$ for the magnetizing case, the shift is in the $-H_B$ direction. Further, the high-$H_C$ end of the ridge, again staying on the $H_C$ axis, gets "stretched" by $-\alpha M_S$, but since $\alpha>0$, this actually means a *compression*.

The appearance of the edge, attached to the ridge at the low-$H_C$ end, can also be understood from arguments analogous to the ones explaining the vertical edge in the demagnetizing case. The first up-flip along each FORC, performed by the lowest coercivity particle $H_K^{Min}$, is biased by $\alpha M(H_R)$, up-flipping at $H_{up}=H_K^{Min}-\alpha M(H_R)$. For the least negative $H_R$ reversal fields $M(H_R)$ is close to $M_S$ and the mean field reduces the up-flip field $H_{up}$ by the largest amount $-\alpha M_S$. For increasingly negative $H_R$ values the magnetization $M(H_R)$ is decreasing, therefore the mean field interactions *reduce $H_{up}$ by a decreasing amount*, in effect shifting the up-flip field $H_{up}$ to increasingly positive values, as shown in FIG. 3e, right panel in the main text.

This shift again makes the first *dM/dH* jumps of the FORCs *unmatched*, but this time by the neighboring FORC with a more negative $H_R$. Moreover, visibly for these magnetizing interactions the value of the *dM/dH* jump changes from positive to zero, making them unmatched. Thus the magnetizing interactions again create the edge but this time with a negative amplitude.

<u>Nearest Neighbor Interactions – demagnetizing model</u>: It is noticed that the measured FORC diagrams exhibit the ridge-and-edge structures, but for the demagnetizing arrays these features are segmented, not smooth. For some of the arrays, the segments are in fact separated from each other.

In an attempt to account for this segmentation of the ridge and edge, in our simulations we chose to augment the mean interaction field with terms explicitly representing the nearest neighbor dipolar interactions, since these are the largest energy terms treated only approximately within the mean field. This extension of the mean field theory can be viewed as including the first terms of a systematic cluster expansion.

As shown in Fig. S1 (c), (f), and (i), the inclusion of nearest neighbor terms in the simulations indeed segment the FORC ridge and edge, verifying our expectations. To understand the effect and phenomenology of the nearest neighbor terms, we now construct a theoretical analysis of the FORC diagram of nearest-neighbor-only models, and then integrate the nearest-neighbor-only model into mean field framework.

As shown in Fig. S3(b), the experimentally relevant two-dimensional demagnetizing arrays with nearest-neighbor-only interactions exhibit three well-defined primary peaks (P1-P3) and three secondary peaks (P4-P6) in the FORC distribution. These features can be directly identified with specific up- and down- switching events. As shown in Fig. 1(a) of the main text, since the coupling of any dipole is strongest to the two nearest neighbors along the minor axis, the interactions can be represented by defining the directions of these three dipoles.

With this convention, and denoting the nearest neighbor interaction field with $H_{n.n.}$, the down-flipping events are:

(D1) positive saturation → checkerboard (↑↑↑ → ↑↓↑: $H_{int}=2H_{n.n.}$),

(D2) checkerboard → negative saturation (↓↑↓ → ↓↓↓: $H_{int}= -2H_{n.n.}$), and

(D3) frustrated checkerboard → frustrated checkerboard (↓↑↑ → ↓↓↑: $H_{int}=0$).

The up-flipping events are:

(U1) checkerboard → positive saturation (↑↓↑ → ↑↑↑: $H_{int}=2H_{n.n.}$),

(U2) negative saturation → checkerboard (↓↓↓ → ↓↑↓: $H_{int}= -2H_{n.n.}$), and

(U3) frustrated checkerboard → frustrated checkerboard (↓↓↑ → ↓↑↑: $H_{int}=0$).

Using the switching conditions discussed in the text, we now describe how each peak in the FORC distribution is generated by a FORC curve, defined by a down-flip and an up-flip. In particular:

- the peak P1 at ($H_C=H_K$, $H_B=+2H_{n.n.}$) is generated by the FORC with the D1 and U1 flips;

- the peak P2 at ($H_C=H_K$, $H_B=-2H_{n.n.}$) is generated by the FORC with the D2 and U2 flips;

- the peak P3 at ($H_C=H_K+2H_{n.n.}$, $H_B=0$) is generated by the FORC with the D2 and U1 flips;

- the peak P4 at ($H_C=H_K+H_{n.n.}$, $H_B=+H_{n.n.}$) is generated by the FORC with the D3 and U1 flips;

- the peak P5 at ($H_C=H_K+H_{n.n.}$, $H_B=-H_{n.n.}$) is generated by the FORC with the D2 and U3 flips; and

- the peak P6 at ($H_C=H_K$, $H_B=0$) is generated by the FORC with the D3 and U3 flips.

Visibly, the amplitudes of peaks P1 and P2 are the strongest. The reason for this is that the state the dipoles flip in-to and out-of when generating the peaks P1/P2 is the lowest energy anti-ferromagnetic state, thus favored by a majority of reversal pathways.

The peak P3 has the next-strongest amplitude. The reason for this is analogous to the above arguments, with the difference that while peaks P1/P2 are formed by flips *from* the saturated states *into* the checkerboard pattern, P3 is generated by high $H_K$ dipoles flipping *from* the checkerboard *into* the saturated states. In sum, these three peaks are well defined because in the demagnetizing systems there exists an energetically favorable intermediate state, the checkerboard state, through which most reversal pathways go through.

In comparison, peaks P4/5/6 have much smaller amplitudes. This is due to the fact that the starting, ending, or both configurations are frustrated, with energies higher than the checkerboard state, and therefore any particular flip sequence is carried out only by a small fraction of the dipoles.

Further, the reversal of a single particle also reduces the interactions on its neighbors, and subsequently they are less likely to reverse, thus nucleating a local checkerboard ordering. Fig. S3(c) illustrates that these locally formed checkerboards tend to organize themselves into larger checkerboard patterns, highlighted by the green boxes. Fig. S3(f) illustrates, that, in contrast, the mean field theory does not capture these local checkerboard-ordering tendencies, and correspondingly, the mean field FORC distribution does not show the formation of localized peaks that could segment the smooth ridge-and-edge structure.

Therefore, to account for the experimentally observed segmenting of the ridge-and-edge structure, the mean field theory and the nearest-neighbor-only frameworks have to be combined. Fig. S3(h)-(i) show the FORC distribution and dipole configuration obtained by simulating the combined model. Both the formation of the ridge-and-edge structure and the 3+3 peaks structure are clearly recognizable in the FORC distribution. To sum-up the demagnetizing considerations, we conclude that both the theoretical analysis and simulations demonstrate that mean field plus nearest neighbor model fully explains all the features in the experimental FORC distributions.

<u>Nearest Neighbor Interactions – magnetizing model:</u> In the experimental systems with magnetizing interactions the FORC distributions was not strongly segmented like it was in the demagnetizing case. We attribute this absence of segmenting to the lack of a lower-energy intermediate state, as the checkerboard state was for the demagnetizing systems. As we show now, in the nearest-neighbor-only model the interactions guide the system from one saturated state directly to the other without getting trapped in any intermediate state. Accordingly, the FORC distribution of these models is a single localized structure, as shown in Fig. S3(k).

In detail, the nearest neighbor magnetizing interactions cause an avalanche-like reversal of chains of dipoles. The effects of the nearest neighbor interactions in these magnetizing systems can be interpreted as follows: the interactions stabilize the parallel ordering between neighbors, aligned along the dipole axis. Therefore, when a dipole flips this *reduces* the stability of the remaining un-flipped dipoles. While the effect is small for the mean-field model, as $\Delta H_{int}=2\alpha M_S/N$, (where N is the total number of dipoles in the system), for nearest neighbor interactions the change in the interactions field is $\Delta H_{int}=2H_{n.n.}$, reducing the total interaction field to zero. This enables the external field to flip the neighboring dipole as well. Consequently, the flip of each dipole greatly de-stabilizes its neighbors and induces them to flip as well. As the neighbors flip, the destabilized front propagates to the next-nearest-neighbors, inducing an avalanche propagating down the chain of neighbors until a local high-coercivity particle ($H_K>2H_{n.n.}+H$) stops the avalanche. The dipoles participating in a single avalanche share the same down-flip and up-flip fields, and therefore contribute to the FORC distribution at a single ($H_C, H_B$) location, which we call the avalanche peak. This single avalanche peak may broaden somewhat for wide coercivity distributions, as the flip field of avalanches is impacted by dipoles with extreme coercivity values.

Fig. S3(n) shows that when the mean field and the nearest-neighbor-only models are combined, since the location of the single avalanche peak overlaps with the mean field ridge, the resulting FORC does not exhibit any segmenting. Rather, it shows a ridge with a well-formed maximum at the location of the avalanche peak.

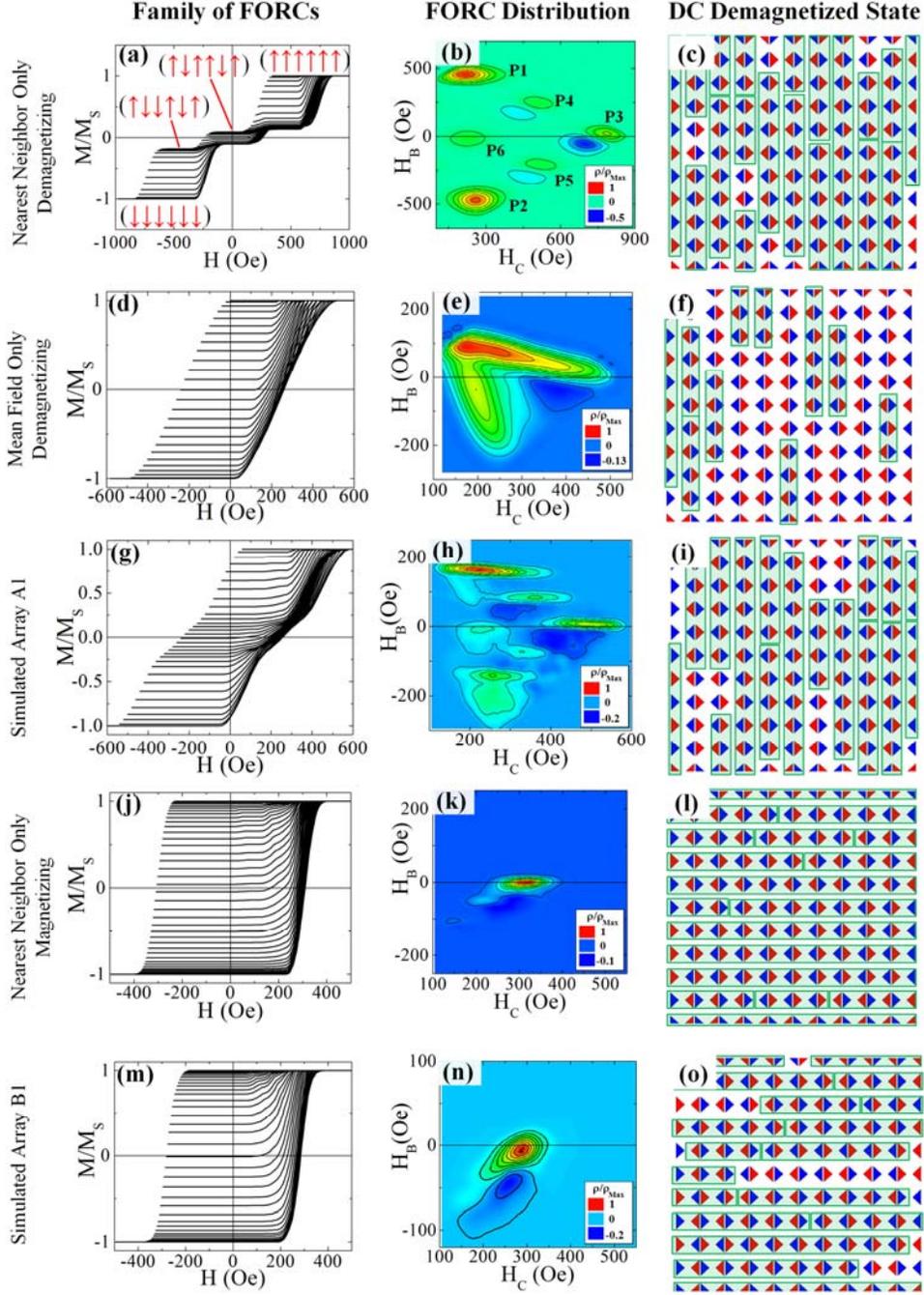

*Fig. S3:* From left to right columns, simulated family of FORCs, FORC distributions and DC-demagnetized remnant states are shown for systems with (a-c) nearest neighbor (n.n.) demagnetizing; (d-f) mean-field (m.f.) demagnetizing; (g-i) combined (m.f.+n.n.) demagnetizing; (j-l) n.n. magnetizing; (m-o) combined (m.f.+n.n.) magnetizing interactions. In the right column, red/blue arrows represent dipole orientations and green boxes highlight local ordering.

9